\documentclass[prd,twocolumn,showpacs,floatfix,amsmath,amssymb,floatfix,nofootinbib]{revtex4}
\usepackage{graphicx,color,dcolumn,booktabs,bm}
\usepackage{longtable,lscape}
\usepackage{txfonts}
\usepackage{overpic}
\usepackage{amssymb}
\usepackage{indentfirst}
\usepackage{feynmf}   
\usepackage{slashed}  
\usepackage{cases}
\usepackage[dvipdfm,  
            pdfborder=001,   
            ]{hyperref}

\begin{document}


\title{Revisiting the production of charmonium plus a light meson at $\overline{\mbox{P}}$ANDA}
\author{Qing-Yong Lin$^{1,2,3}$}\email{qylin@impcas.ac.cn}
\author{Hu-Shan Xu$^{1,2}$}
\affiliation{
$^1$Institute of Modern Physics, Chinese Academy of Sciences, Lanzhou 730000, China\\
$^2$Research Center for Hadron and CSR Physics, Lanzhou University and Institute of Modern Physics of CAS, Lanzhou 730000, China\\
$^3$Graduate University of Chinese Academy of Sciences, Beijing, 100049, China}
\author{Xiang Liu$^{1,2}$\footnote{corresponding author}}\email{xiangliu@lzu.edu.cn}
\affiliation{$^1$School of Physical Science and Technology, Lanzhou University, Lanzhou 730000, China\\
$^2$Research Center for Hadron and CSR Physics, Lanzhou University and Institute of Modern Physics of CAS, Lanzhou 730000, China}

\date{\today}

\begin{abstract}
In this work, we calculate the total cross sections and the center-of-mass frame angular distributions of the charmonium production plus a light meson by the low energy $p\bar{p}$ interaction. The results of $p\bar{p}\to \pi^0 \Psi$ with and without form factor (FF) indicate that the FF contribution in the calculation cannot be ignored. The obtained cross section of $p\bar{p}\to \pi^0 J/\psi$ with FF can fit the E760 data well. We also predict the total cross sections and the center-of-mass frame angular distributions of $p\bar{p}\to \omega \Psi$, which show that these physical quantities are dependent on Pauli ($g_\omega$) and Dirac ($\kappa_\omega$) coupling constants of the $pp\omega$ interaction. Thus, $p\bar{p}\to \omega \Psi$ can be as the ideal channel to test the different theoretical values of $g_\omega$ and $\kappa_\omega$. Applying the formulae of $p\bar{p}\to \pi^0 \Psi$ and $p\bar{p}\to \omega \Psi$, we predict the total cross sections of the $p\bar{p}\to \eta \Psi$ and $p\bar{p}\to \rho \Psi$ reactions. Our results show a common behavior of the charmonium production with a light meson by the $p\bar{p}$ interaction, where the total cross section of the $\eta_c$ production is the largest one among all discussed processes. The above observations can be directly tested at the forthcoming $\overline{\mbox{P}}$ANDA experiment.
\end{abstract}

\maketitle


\section{Introduction}\label{sec1}

Since the discovery of $J/\psi$ in 1974 \cite{Aubert:1974js,Augustin:1974xw}, more and more charmonia have been reported by experiment \cite{Nakamura:2010zzi}, which provide an ideal platform to improve our understanding of non-perturbative Quantum Chromodynamics (QCD) dynamics. In the past decade, the observations of a series of charmonium-like states named as $XYZ$ state have stimulated extensive interest in studying higher charmonia for both theorists and experimentalists. As one of the forthcoming experiments relevant to the study of hadron physics, Antiproton Annihilations at Darmstadt ($\overline{\mbox{P}}$ANDA) experiment at the Facility for Antiproton and Ion Research (FAIR) can serve as the investigation of charmonium, which is also one of the main physical aims of $\overline{\mbox{P}}$ANDA \cite{Lutz:2009ff}. Thus, theoretically studying the charmonium production by the low energy $p\bar{p}$ interaction
becomes an important and interesting research work, which can give valuable suggestions for the forthcoming $\overline{\mbox{P}}$ANDA experiment.

In Ref. \cite{Gaillard:1982zm}, Gaillard and Maiani firstly calculated the differential cross section of the charmonium production accompanied by a soft pion in the low energy $p\bar{p}$ interaction, where two hadron-level diagrams were introduced by the Born approximation. They indicated that the corresponding cross section is proportional to the partial decay width of charmonium decay into $p\bar{p}$ \cite{Gaillard:1982zm}.  The authors of Ref. \cite{Lundborg:2005am} further studied the cross sections of the charmonium ($\Psi$) production plus a light meson ($m$) by the $p\bar{p}\to \Psi+m$ processes, which can be related to the measured partial decay width of charmonium decay into $p\bar{p}m$ \cite{Lundborg:2005am}. By this approach, the cross sections of $p\bar{p}\to J/\psi (\psi^\prime) m$ ($m=\pi^0, \eta, \rho^0, \omega, \eta^\prime, \phi$) are estimated. Among these predicted cross sections, the cross section of the $p\bar{p}\to \pi^0 J/\psi$ process can reach up to 300 pb when taking $\sqrt{s}=3$ GeV, which is close to the E760 data taking $\sqrt{s}=3.5\sim 3.6$ GeV \cite{Armstrong:1992ae}. Later, Barnes and Li developed the initial state light meson emission model, which was applied to study the near threshold associated charmonium production process $p\bar{p}\to \pi^0 \Psi$, where $\Psi$ denotes $\eta_c$, $J/\psi$, $\psi^\prime$, $\chi_{c0}$,  $\chi_{c1}$ \cite{Barnes:2006ck}. By the initial state light meson emission model, they calculated the differential cross section and total cross sections of $p\bar{p}\to \pi^0 \Psi$ processes. In addition, $\langle d\sigma/d\Omega\rangle$, the center-of-mass frame unpolarized angular distribution, was predicted for $p\bar{p}\to \pi^0 \Psi$ \cite{Barnes:2006ck}. In Ref. \cite{Barnes:2007ub}, Barnes {\it et al.} further indicated that the cross section of $p\bar{p}\to \pi^0 J/\psi$ near threshold may be affected by the Pauli $J/\psi p\bar{p}$ coupling, which will be an interesting research topic in $\overline{\mbox{P}}$ANDA \cite{Barnes:2007ub}.

By the initial state light meson emission model \cite{Barnes:2006ck}, $p\bar{p}\to \pi^0 \Psi$ occurs via a proton exchange between $p$ and $\bar{p}$. In Ref. \cite{Barnes:2006ck}, authors treated the $ppm$ and $J/\psi p\bar{p}$ couplings as the point-like interaction vertices. However, in reality we should consider the structure effect of the $ppm$ and $J/\psi p\bar{p}$ interactions. Thus, for reflecting such structure effect, the form factor should be introduced in the $ppm$ and $J/\psi p\bar{p}$ interaction vertices, which was listed as one of the future developments of the initial state light meson emission model \cite{Barnes:2006ck}. Along this way, in this work we revisit the production of charmonium plus a light meson in the low energy $p\bar{p}$ interaction by considering the contribution of form factor (FF) to these processes. The comparison of the results with and without including FF in the calculation can reveal the difference under two cases, which will be tested at the forthcoming $\overline{\mbox{P}}$ANDA experiment. By this study, we can not only extract the physical picture depicting the production of charmonium plus a light meson in the low energy $p\bar{p}$, but also learn what kind of FF to be suitable to describe the structure effect of the $ppm$ and $J/\psi p\bar{p}$ interactions. These processes discussed in this work also include the charmonium production with a light vector meson ($\omega$, $\rho$), where the $pp\omega$ or $pp\rho$ interaction relates to both Dirac and Pauli couplings. At present, the coupling constants of Dirac and Pauli couplings of $pp\omega$ and $pp\rho$ are determined by some theoretical groups by different processes and different models \cite{Barnes:2010yb, Downum:2006re, Cottingham:1973wt, Lacombe:1980dr, Nagels:1978sc, Machleidt:2000ge, Sato:1996gk, Mergell:1995bf, Zhu:1999kva}. Thus, we adopt these determined coupling constants in our calculation, which includes the total cross sections, the differential cross sections $\langle d\sigma/d\Omega\rangle$. These studies can serve as further experimental test of these coupling constants by the charmonium production with a light vector meson at $\overline{\mbox{P}}$ANDA.

This work is organized as follows. After introduction, we present the calculation of the production of charmonium plus a light meson in the low energy $p\bar{p}$ interaction. In Sec. \ref{sec3}, the numerical results are given. The last section is the discussion and conclusion.

\section{The production of charmonium}\label{sec2}

\begin{figure}[htb]
\begin{center}
\scalebox{0.45}{\includegraphics{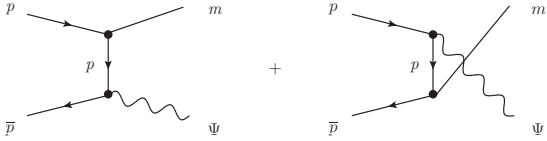}}
\caption{The diagrams describing the production charmonium plus a light meson by the $p\bar p$ interaction. Here, $\Psi$ denotes charmonium ($\eta_c$, $J/\psi$, $\psi'$, $\chi_{c0}$, $\chi_{c1}$) while $m$ is light meson ($\pi^0$, $\eta$, $\rho$, $\omega$).
\label{fig:FeynmanDiagram}}
\end{center}
\end{figure}

As depicted by Fig. \ref{fig:FeynmanDiagram}, the charmonium ($\Psi$) production plus a light meson ($m$) by the low energy $p\bar{p}$ interaction can occur via the transition of $p\bar{p}$ into $\Psi+m$ by exchanging a proton. Thus, there exist two hadron-level diagrams \cite{Gaillard:1982zm,Barnes:2006ck} shown in Fig. \ref{fig:FeynmanDiagram} if only considering the tree-level contributions.

When deducing the corresponding production amplitude, we use effective Lagrangian approach. The interaction vertices of $ppm$ and $pp\Psi$ include
\begin{numcases}
{ \mathcal{L}_{ppm}= }
-ig_{NN\pi}\bar \phi \gamma_5 {\mbox{\boldmath $\tau$}}\cdot {\mbox{\boldmath $\pi$}}  \phi, &for $m=\pi^0$ \nonumber \\
-g_{NN\omega}\left(\bar \phi\gamma_\mu\phi\omega_\mu -\frac{\kappa_\omega}{4m_p}\bar \phi\sigma_{\mu\nu}\phi F_{\mu\nu}\right),& for $m=\omega$ \nonumber
\end{numcases}
and
\begin{numcases}
{\mathcal{L}_{pp\Psi}= }
-ig_{NN\eta_c}\bar \phi\gamma_5\phi\eta_c,&for $\Psi=\eta_c$ \nonumber\\
-g_{NN\chi_{c0}}\bar \phi \phi\chi_{c0},& for $\Psi=\chi_{c0}$ \nonumber\\
-g_{NNJ/\psi(\psi')}\bar \phi\gamma_\mu \phi\psi_\mu,& for $\Psi=J/\psi(\psi^\prime)$\nonumber\\
-g_{NN\chi_{c1}}\bar \phi \gamma_\mu\gamma_5 \phi\chi_{c1}^{\mu},& for $\Psi=\chi_{c1}$ \nonumber
\end{numcases}
where $\phi$ denotes the field of nucleon and the ${\mbox{\boldmath $\tau$}}$ is Pauli matrix. $\sigma_{\mu\nu}=i[\gamma_\mu,\gamma_\nu]/2$ and $F_{\mu\nu}=\partial_\mu \omega_\nu - \partial_\nu \omega_\mu$.
For the $pp\omega$ interaction, there exists two independent coupling constants, {\it i.e}, $g_{\omega}\equiv g_{NN\omega}$ and $\kappa_\omega$ corresponding to Dirac and Pauli terms respectively.

In the following, we illustrate the calculation of the charmonium production.
Without introducing FF in each of interaction vertices in Fig. \ref{fig:FeynmanDiagram}, the resulting amplitude of $p\bar p \to \pi^0\Psi$ is \cite{Barnes:2006ck}
\begin{eqnarray}\label{eq:pi_amplitude}
    \mathcal{M}_{p\bar p \to \pi^0\Psi}
    &=& g_{\pi} g_{\Psi}{\bar v}_{\bar{p}}(p_2,s_2)\bigg(\Gamma_1 \frac{(\slashed{p}_1-\slashed{k}+m_p)}{(t-m_p^2)}\gamma^5 \nonumber\\ &&+ \gamma^5 \frac{(\slashed{k}- {\slashed{p}}_2+m_p)}{(u-m_p^2)}\Gamma_1\bigg) u_{p}(p_1,s_1),\label{pi}
\end{eqnarray}
where $m_p$ is the mass of proton. $p_1$, $p_2$ and $k$ are the four momenta of proton, antiproton and the emitted light meson. $s$, $t$ and $u$ are the Mandelstam variables with the definitions $s=(p_1+p_2)^2$, $t=(p_1-k)^2$ and $u=(k-p_2)^2$. $g_\pi\equiv g_{pp\pi}$ and $g_{_\Psi}\equiv g_{p\bar p\Psi}$ denote the
coupling constants of the $pp\pi$ and $pp\Psi$ interactions, respectively. $\bar{\nu}_{\bar{p}}$ and $u_{p}$ denote the spinors of antiproton and proton, respectively. In Eq. (\ref{pi}), $\Gamma_1$ is defined as $\gamma_5$, $-i$, $-i\gamma_\mu \epsilon_{J/\psi}^\mu$ and $-i\gamma_\mu\gamma_5\epsilon_{\chi_{c1}}^\mu$ corresponding to $p\bar p\to \pi^0\Psi$ processes with $\Psi$ taken as $\eta_c$, $\chi_{c0}$, $J/\psi(\psi^\prime)$ and $\chi_{c1}$, respectively.

For the $p\bar{p}\to \omega \Psi$ process discussed here, its production amplitude can be written as
\begin{eqnarray}\label{eq:omega_amplitude}
    &&\mathcal{M}_{p\bar{p}\to \omega \Psi}\nonumber\\&&= g_{\omega} g_{_\Psi}{\bar v}_{\bar p}(p_2, s_2)\left(\Gamma_2 \frac{1}{\slashed{p}_1-\slashed{k}-m_p}\gamma^{\mu} + \gamma^{\mu} \frac{1}{\slashed{k}-\slashed{p}_2-m_p}\Gamma_2\right)
    \nonumber\\
    &&\quad  \times u_{p}(p_1, s_1) \epsilon_{\mu}^{\ast}+ i\frac{\kappa_{\omega} g_{\omega} g_{\Psi}}{2 m_p} {\bar v}_{\bar p}(p_2, s_2) \left(\Gamma_2 \frac{1}{\slashed{p}_1-\slashed{k}-m_p}\right.
    \nonumber\\
    && \left. \quad \times\sigma^{\mu\nu} k_{\nu}+ \sigma^{\mu\nu} k_{\nu}\frac{1}{\slashed{k}-\slashed{p}_2-m_p}\Gamma_2\right) u_{p}(p_1, s_1)\epsilon_{\mu}^{\ast}
    \nonumber\\
    &&= g_{\omega} g_{_\Psi}{\bar v}_{\bar p}(p_2, s_2)\left(\Gamma_2 \frac{(\slashed{p_1}-\slashed{k}+m_p)}{(t-m_p^2)}\gamma^{\mu}\right.
    \nonumber\\
    && \left.\quad + \gamma^{\mu} \frac{(\slashed{k}-\slashed{p}_2+m_p)}{(u-m_p^2)}\Gamma_2\right)
     u_{p}(p_1, s_1)\epsilon_{\mu}^{\ast} \nonumber\\&&\quad+ i\frac{\kappa_{\omega} g_{\omega} g_{\Psi}}{2 m_p} {\bar v}_{\bar p}(p_2, s_2) \left(\Gamma_2 \frac{(\slashed{p}_1-\slashed{k}+m_p)}{(t-m_p^2)}\sigma^{\mu\nu} k_{\nu}\right.
    \nonumber\\
    && \left.\quad + \sigma^{\mu\nu} k_{\nu}\frac{(\slashed{k}-\slashed{p}_2+m_p)}{(u-m_p^2)}\Gamma_2\right) u_{p}(p_1, s_1)\epsilon_{\mu}^{\ast},\label{omega}
\end{eqnarray}
where $\epsilon_{\mu}^{\ast}$ is the polarization four-vector of the emitted $\omega$ meson. The definition of $\Gamma_2$ is the same as that of $\Gamma_1$ in calculating $p\bar{p}\to \pi^0 \Psi$. 

Just discussed above, the FF contribution to $p\bar{p}\to \pi^0 \Psi$ and $p\bar{p}\to \omega \Psi$ is not included in Eqs. (\ref{pi}) and (\ref{omega}). Since the $ppm$ and $pp\Psi$ vertices are not the point-like  interactions, we need to introduce FF in each of interaction vertices, where the FF not only reflects the structure effect of the interaction vertex but also plays an important role to compensate the off-shell effect of the exchanged proton. For comparing the results with and without FF, we calculate the $p\bar{p}\to \pi^0 \Psi$ and $p\bar{p}\to \omega \Psi$ processes considering the FF contribution, where the corresponding amplitudes can be expressed as
\begin{eqnarray}\label{eq:pi_amff}
    \mathcal{M}^{FF}_{\bar p p\rightarrow \pi^0 \Psi}
    &=& g_{\pi} g_{_\Psi}{\bar v}_{\bar p}(p_2, s_2)\bigg[\Gamma_1 \frac{(\slashed{p}_1-\slashed{k}+m_p)}{(t-m_p^2)}\gamma^5\mathcal{F}^2(q_{t}^2)
    \nonumber\\
    &&  +\gamma^5 \frac{(\slashed{k}-\slashed{p}_2+m_p)}{(u-m_p^2)}\Gamma_1\mathcal{F}^2(q_{u}^2)\bigg] u_{p}(p_1, s_1)
\end{eqnarray}
and
\begin{eqnarray}\label{eq:omega_amff}
    &&\mathcal{M}^{FF}_{\bar p p\rightarrow \omega \Psi}\nonumber\\
    &&= g_{\omega} g_{_\Psi}{\bar v}_{\bar p}(p_2, s_2)\left[\Gamma_2 \frac{(\slashed{p}_1-\slashed{k}+m_p)}{(t-m_p^2)}\gamma^{\mu}\mathcal{F}^2(q_{t}^2) + \gamma^{\mu}
    \right.
    \nonumber\\
    && \left. \quad\times\frac{(\slashed{k}-\slashed{p}_2+m_p)}{(u-m_p^2)} \Gamma_2\mathcal{F}^2(q_{u}^2)\right] u_{p}(p_1, s_1) \epsilon_{\mu}^{\ast} + i\frac{\kappa_{\omega} g_{\omega} g_{_\Psi}}{2 m_p}
    \nonumber\\
    &&  \quad\times{\bar v}_{\bar p}(p_2, s_2)\bigg[\Gamma_2 \frac{(\slashed{p}_1-\slashed{k}+m_p)}{(t-m_p^2)}\sigma^{\mu\nu}k_{\nu}\mathcal{F}^2(q_{t}^2) + \sigma^{\mu\nu} k_{\nu}
    \nonumber\\
    && \quad\times\frac{(\slashed{k}-\slashed{p}_2+m_p)}{(u-m_p^2)}\Gamma_2\mathcal{F}^2(q_{u}^2)\bigg] u_{p}(p_1, s_1)\epsilon_{\mu}^{\ast},
\end{eqnarray}
where $\mathcal{F}^2(q_i^2)$ denotes the introduced monopole FF with the definition ${\cal F}(q_i^2)=(\Lambda^2-m_i^2)/(\Lambda^2-q_i^2)$. $m_i$ and $q_i$ are the mass and the four-momentum of the exchanged proton, respectively. $\Lambda$ is a free parameter, which is expected to be around 1 GeV. In the next section, we will discuss the value of $\Lambda$ adopted in our calculation. In the above expressions, $q_t$ and $q_u$ denote the four-momenta of the exchanged protons of t-channel and u-channel of the $p\bar{p}\to m \Psi$ process just shown in Fig. \ref{fig:FeynmanDiagram}, respectively. In addition, the superscript $FF$ is introduced for distinguishing the amplitudes with and without FF. As indicated in Ref. \cite{Lansberg:2010mf}, the $p\bar{p}\to m\Psi$ processes may include transition distribution amplitudes \cite{Pire:2005ax}, which generalize the form factors that we include in the $ppm$ vertex \cite{Lansberg:2010mf}.

We notice that the amplitudes listed in Eqs. (\ref{pi}) and (\ref{omega}) are indeed transverse. When introducing FF in these amplitudes, we cannot make these amplitudes keep transverse.
Of course, finding a more suitable form of FF is an interesting research topic, where this FF not only reflects the realistic physical picture but also can make the corresponding amplitudes still be transverse. In this work, we still choose the monopole FF in our calculation for reflecting the realistic physical picture, and estimate the production of charmonium by the $p\bar{p}$ interaction processes.

The general differential cross section of $p\bar{p}\to m \Psi$ is given by \cite{Nakamura:2010zzi}
\begin{eqnarray}\label{eq:dt}
    \frac{d\sigma}{dt} = \frac{1}{64\,\pi\, s}\frac{1}{|{\bm{p}}_{1cm}|^2}\overline{|\mathcal{M}|}^2 ,
\end{eqnarray}
where ${\bm{p}}_{1cm}$ is the three-momentum of proton in the center-of-mass frame of $p\bar{p}$. The overline indicates the average over the polarizations
of $p/\bar{p}$ in the initial state and the sum over
the polarization of $m/\Psi$ in the final state.

With these obtained amplitudes listed in Eqs. (\ref{eq:pi_amplitude})-(\ref{eq:omega_amff}), we finally get
the expressions of the differential cross sections of the $p\bar{p}\to \omega\Psi$ processes without including the FF contribution, which are collected in appendix \ref{Appendix_A}. The corresponding total cross sections of $p\bar{p}\to \omega\Psi$ are shown in appendix \ref{Appendix_B}. We also confirm the deductions of  the differential and total cross sections of $p\bar{p}\to \pi^0\Psi$ in Ref. \cite{Barnes:2006ck}. Since the formulae of the differential and total cross sections of $p\bar{p}\to \pi^0\Psi$ and $p\bar{p}\to \omega\Psi$ with the FF contribution are very complicated, we do not show their concrete expressions in detail, but directly apply their formulae to the numerical calculation.

\section{ Numerical results }\label{sec3}

Before presenting the numerical result, we first introduce the coupling constants adopted in our calculation.
For the processes $p\bar p \rightarrow \pi^0\Psi$, the coupling constant of the $pp\pi$ interaction was given in many theoretical works, where we take  $g_{NN\pi}=13.5$ \cite{Lin:1999ve}. The coupling constants of charmonium interacting with nucleons are not well established at present. Thus, in this work we adopted the same values as those in Ref. \cite{Barnes:2006ck}, where these coupling constants are $g_{p\bar p \eta_c} = (19.0 \pm 3.2)\times 10^{-3}$, $ g_{p\bar p J/\psi} = (1.62 \pm 0.03)\times 10^{-3} $, $g_{p\bar p \psi'} =(0.97 \pm 0.04)\times 10^{-3} $, $g_{p\bar p \chi_{c0}} =(5.42 \pm 0.37)\times 10^{-3}$, $g_{p\bar p \chi_{c1}} =(1.03 \pm 0.07)\times 10^{-3} $, which were estimated by the measured partial width of $\Psi\to p\bar{p}$ \cite{Nakamura:2010zzi}.

For the discussed processes $p\bar p \rightarrow \omega\Psi$, there are two strong coupling constants $g_\omega$ and $\kappa_\omega$ for the $pp\omega$ vertex. At present, different models and different approaches gave various theoretical values for these two coupling constants, which are listed in Table. \ref{tab:omega_couplings}.
\renewcommand{\arraystretch}{1.3}
\begin{table}[htb]
\caption{The estimated values of the coupling constants $g_\omega$ and $\kappa_\omega$.
\label{tab:omega_couplings}}
\begin{tabular}{c|c|c}
\toprule[1pt]
Mechanism/Model & $g_\omega$ & $\kappa_\omega$ \\\midrule[0.5pt]

Paris \cite{Cottingham:1973wt, Lacombe:1980dr} & 12.2 & $-0.12$ \\
Nijmegen \cite{Nagels:1978sc} & 12.5 & $+0.66$ \\
Bonn \cite{Machleidt:2000ge} & 15.9 & 0 \\\midrule[0.5pt]

Pion photoproduction \cite{Sato:1996gk} & $7-10.5$ & 0 \\\midrule[0.5pt]

Nucleon EM form factors \cite{Mergell:1995bf} & $20.86\pm0.25$ & $-0.16\pm0.01$ \\\midrule[0.5pt]

QCD sum rule \cite{Zhu:1999kva} & $18\pm8$ & $0.8\pm0.4$ \\\midrule[0.5pt]

$^3P_0$ quark model \cite{Downum:2006re} & $-$ & $-3/2$ \\\midrule[0.5pt]

& $23\pm3$ & 0
\\
Light meson emission model \cite{Barnes:2010yb} & &  \\
& $14.6\pm2.0$ & $-3/2$
\\
\bottomrule[1pt]
\end{tabular}
\end{table}

In addition, the masses of the hadrons involved in our calculation include $m_{\pi^0} = 135.0$ MeV, $m_{\omega} = 782.7$ MeV, $m_{p} = 938.3$ MeV, $m_{\eta_c} = 2980.3$ MeV, $m_{J/\psi} = 3096.9$ MeV, $m_{\chi_{c0}} = 3414.8$ MeV, $m_{\chi_{c1}} = 3510.7$ MeV and $m_{\psi^\prime} = 3686.1$ MeV, which are from Particle Data Group \cite{Nakamura:2010zzi}.

\subsection{The total cross sections and angular distributions of $p\bar p \to \pi^0\Psi$}

\begin{center}
\begin{figure}[hbt]
\scalebox{0.55}{\includegraphics{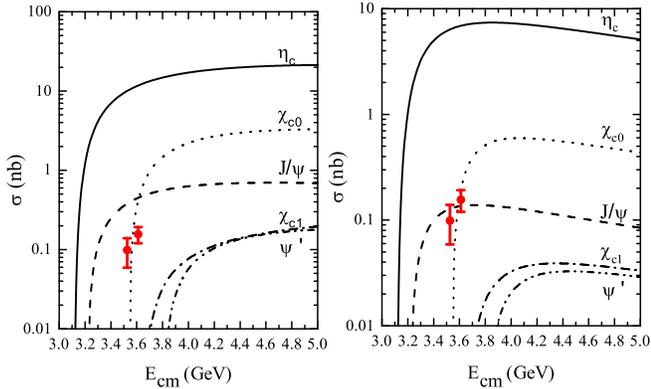}}
\caption{(color online). The obtained total cross section of $p\bar p \rightarrow \pi^0\Psi$ and the comparison of the experimental and theoretical results. Here, the red points with errors are the experimental measurement from E760 \cite{Armstrong:1992ae}. The left-hand and right-hand diagrams correspond to the theoretical results without and with the FF contribution to $p\bar p\to \pi^0\Psi$, respectively.}
\label{fig:piCStotal}
\end{figure}
\end{center}

Firstly, in this work we confirm the results of the total cross sections of $p\bar p \to \pi^0\Psi$ in Ref. \cite{Barnes:2006ck}, where the $pp\Psi$ and $pp\pi$ vertices are as point-like structures. Just shown in the left-hand diagram of Fig. \ref{fig:piCStotal}, the dependence of the total cross sections of $p\bar p \to \pi^0 \Psi$ on the center-of-mass energy $E_{cm}$ is presented with a comparison with the experimental data from the E760 experiment \cite{Armstrong:1992ae}, where the cross section of $p\bar p \rightarrow \pi^0J/\psi$ near 3.5 GeV was measured at Fermilab. The theoretical result of $p\bar p \to \pi^0 J/\psi$ taking $E_{cm}=3.5$ GeV is larger than the experimental value obviously.

In Sec. \ref{sec2}, we investigate the structure effect of the $pp\Psi$ and $pp\omega$ vertices on the total cross sections of $p\bar p \to \pi^0\Psi$, which is presented in the right-hand diagram in Fig. \ref{fig:piCStotal}. When taking $\Lambda=1.9$ GeV, we obtain the total cross section of $p\bar p \rightarrow \pi^0J/\psi$ at $E_{cm}=3.5$ GeV consistent with the experimental data, which indicates that the FF contribution cannot be ignored in studying $p\bar p \to \pi^0\Psi$. Adopting the same $\Lambda$ value, we also obtain the total cross sections of other charmonium productions, which are listed in the right-hand diagram of Fig. \ref{fig:piCStotal}. By checking the results with and without considering the FF contribution, we find that there exist differences, where the obtained total cross sections of $p\bar p \to \pi^0\Psi$ with FF are suppressed compared with those without FF, which are shown in Fig. \ref{fig:piCStotal}. In addition, with increasing $E_{cm}$, the total cross section with FF goes down after reaching its maximum, while the total cross section without FF is goes up continuously. The forthcoming $\overline{\mbox{P}}$ANDA experiment can test these theoretical results.

\begin{center}
\begin{figure}[htb]
\scalebox{0.55}{\includegraphics[]{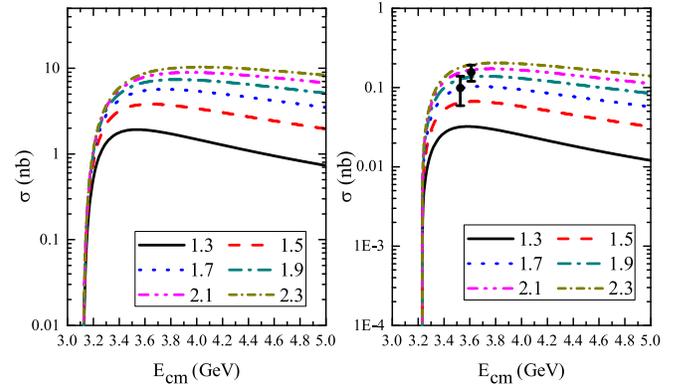}}
\caption{(color online). The total cross sections of $p\bar p \to \pi^0\eta_c$ (left-hand) and $p\bar p \to \pi^0 J/\psi$ (right-hand) with different $\Lambda$ values. Here, the black points with error bars are the E760 data \cite{Armstrong:1992ae}.}
\label{fig:piCSff}
\end{figure}
\end{center}

With $p\bar p \to \pi^0\eta_c$ and $p\bar p \to \pi^0 J/\psi$ as example, we also present the variation of their total cross sections with different $\Lambda$ values as shown in Fig. \ref{fig:piCSff}, where we take several typical values in the range of $\Lambda=1.3\sim 2.3$ with step of 0.2 GeV. These results show that the total cross sections of $p\bar p \to \pi^0\eta_c$ and $p\bar p \to \pi^0 J/\psi$ depend on the value of $\Lambda$.

\begin{center}
\begin{figure}[htb]
\scalebox{0.7}{\includegraphics{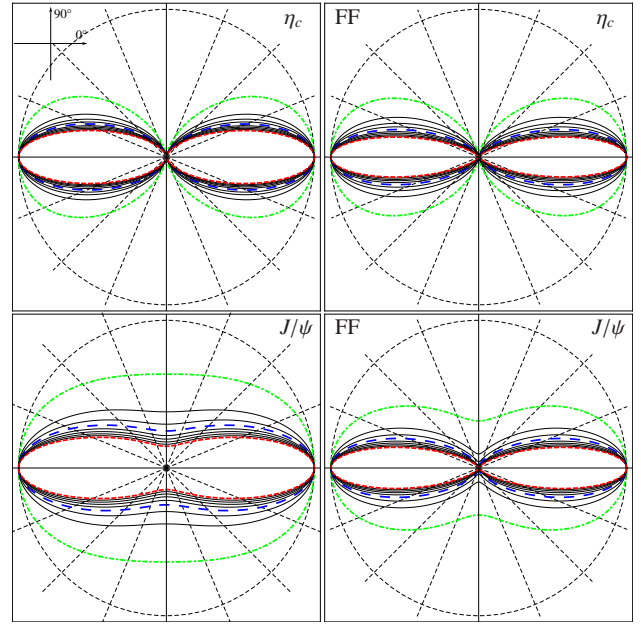}}
\caption{(color online). The center-of-mass frame angular distribution $d\sigma / d\Omega$ of $p\bar p \to \pi^0\eta_c$ and $p\bar p \to \pi^0 J/\psi$. Here, the results are given by taking the range of $E_{cm}=3.2-5.0$ GeV or $3.4-5.0$ GeV with step of 0.2 GeV for the $\eta_c$ or $J/\psi$ production.
The diagrams in the first column are the results without FF \cite{Barnes:2006ck} while the remaining diagrams
are the results with FF, where we take $\Lambda=1.9$ GeV. The results at the lower and upper limits of $E_{cm}$ range are highlighted with dot-dashed green and short-dashed red lines respectively, while long-dashed blue lines are the result at $E_{cm}=4$ GeV. All results are normalized to the forward intensity. }
\label{fig:piAD}
\end{figure}
\end{center}

\begin{center}
\begin{figure}[htb]
\scalebox{0.55}{\includegraphics{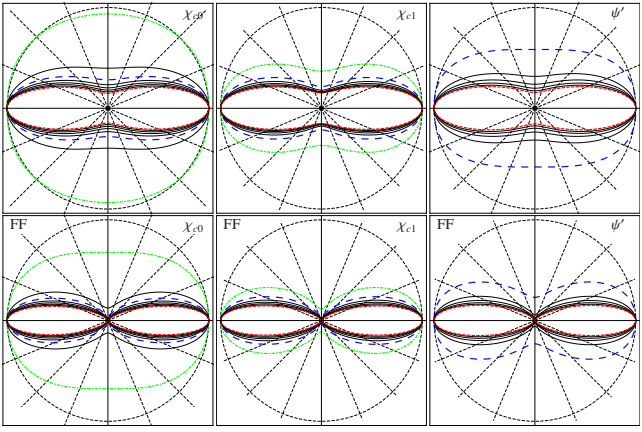}}
\caption{(color online). The center-of-mass frame angular distribution $d\sigma / d\Omega$ of $p\bar p \to \pi^0\chi_{c0}$, $p\bar p \to \pi^0 \chi_{c1}$ and $p\bar p \to \pi^0 \psi^\prime$ corresponding to $E_{cm}=3.6-5.0$ GeV, $3.8-5.0$ GeV, $4.0-5.0$ GeV with step of 0.2 GeV, respectively. Here, the results without and with FF correspond to these diagrams in the first and the second rows, respectively. The style of the highlighted lines is arranged in the same way as in Fig. \ref{fig:piAD}.}
\label{fig:piAD1}
\end{figure}
\end{center}

Besides providing the information of total cross section, we also give the result of the center-of-mass unpolarized angular distributions of $p\bar p \to \pi^0\eta_c$ and $p\bar p \to \pi^0 J/\psi$ just suggested in Ref. \cite{Barnes:2006ck}. Considering the FF contribution, the corresponding angular distributions $d\sigma/d\Omega$ are shown in the second column of Fig. \ref{fig:piAD}, where the results without FF are also listed. In addition, the center-of-mass unpolarized angular distributions of
$p\bar p \to \pi^0\chi_{c0}$, $p\bar p \to \pi^0 \chi_{c1}$ and $p\bar p \to \pi^0 \psi^\prime$ are listed in Fig. \ref{fig:piAD1}. We can find that the angular distributions of $p\bar p \to \pi^0\Psi$ with and without FF are anisotropic just shown in Figs. \ref{fig:piAD} and \ref{fig:piAD1}. Another common peculiarity of these results in Figs. \ref{fig:piAD} and \ref{fig:piAD1} is that these obtained distributions become forward- and backward-peaked with increasing $E_{cm}$, which is consistent with the conclusion in Ref. \cite{Barnes:2006ck}.

The comparison of the results in Figs. \ref{fig:piAD} and \ref{fig:piAD1} also indicates the different behaviors of the center-of-mass frame angular distributions $d\sigma / d\Omega$ with and without FF, especially for $p\bar p \to \pi^0 \psi^{\prime}$, $p\bar p \to \pi^0\chi_{c0}$ and $p\bar p \to \pi^0\chi_{c1}$. For $p\bar p \to \pi^0 \eta_c$, its $d\sigma / d\Omega$ distribution with FF become more forward- and backward-peaked than that without FF when taking the same $E_{cm}$ value. There obviously exists a node in its differential cross sections with and without FF when taking  $\theta=90^\circ$ in the center-of-mass frame or $t=u$, which is totally different from the corresponding result without FF, where $\theta$ is defined as the angle between the light meson and proton. In fact, this behavior is also supported by analyzing the detailed analytic expressions of this process.
For $p\bar p \to \pi^0 J/\psi$, we notice that its differential cross section with FF
is close to 0 when taking $E_{cm}=5$ GeV and $\theta=90^\circ$ in the center-of-mass frame. The situations of $p\bar p \to \pi^0 \psi^{\prime}$, $p\bar p \to \pi^0\chi_{c0}$ and $p\bar p \to \pi^0\chi_{c1}$ are similar to that of $p\bar p \to \pi^0 J/\psi$. These results further indicate that the FF contribution should be considered in studying the charmonium production plus a light meson at the low energy $p\bar{p}$ interaction.

The above investigation of the center-of-mass unpolarized angular distributions of $p\bar p \to \pi^0\Psi$ can provide valuable information to the design of the $\overline{\mbox{P}}$ANDA detector and the analysis of the experimental data. What is more important is that further experiment can test the calculated total cross sections and the $d\sigma / d\Omega$ distributions of $p\bar p \to \pi^0\Psi$.

\subsection{The obtained result for the $p\bar p \to \omega \Psi$ reactions}

\begin{center}
\begin{figure}[htb]
\scalebox{0.75}{\includegraphics{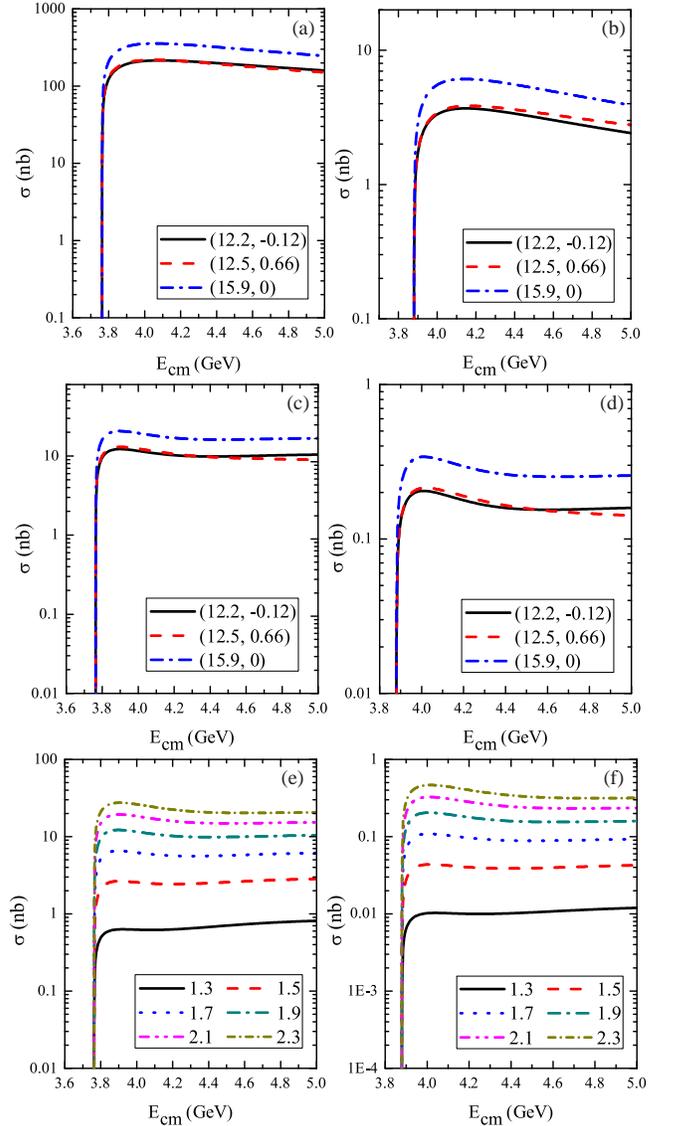}}
\caption{The total cross sections of $p \bar p \to \omega\eta_c$ (left column) and $p \bar p \to \omega J/\psi$ (right column) dependent on $E_{cm}$.
The diagrams in the first row are the result without FF, while the rest diagrams are the result with FF. Here, we take three different combinations of $g_\omega$ and $\kappa_\omega$ to illustrate the total cross sections dependent on $(g_\omega,\kappa_\omega)$ just shown in (a)-(d).
Among these four diagrams, diagrams (c) and (d) are obtained by taking $\Lambda=1.9$ GeV. Diagrams (e) and (f) show the total cross sections with different $\Lambda$ values and the typical value $(g_\omega, \kappa_\omega)$=(12.2, -0.12).}
\label{fig:omegaCS_CC}
\end{figure}
\end{center}

\begin{center}
\begin{figure}[htb]
\scalebox{0.75}{\includegraphics{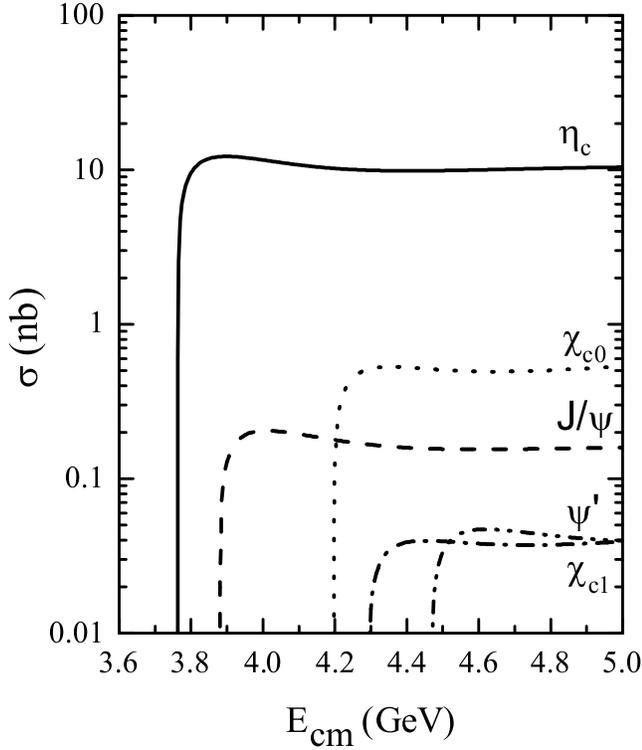}}
\caption{The predicted total cross sections of $p \bar p \to \omega\Psi$ corresponding to typical values $\Lambda=1.9$ GeV and $(g_{\omega}, \kappa_{\omega}) = (12.2, -0.12)$.}
\label{fig:omegaCStotal}
\end{figure}
\end{center}

\begin{center}
\begin{figure}[htb]
\scalebox{0.6}{\includegraphics{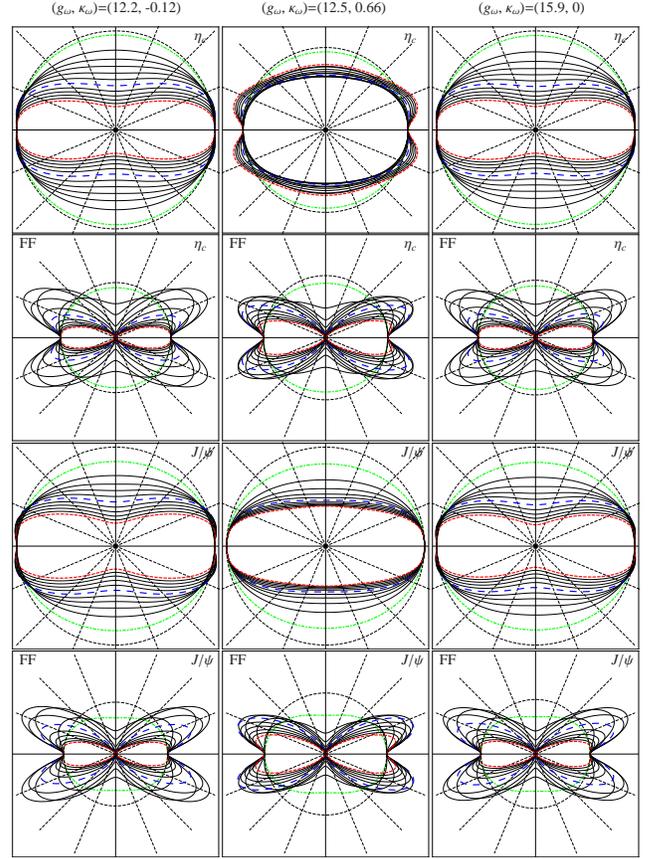}}
\caption{(color online). The obtained center-of-mass frame angular distribution $d\sigma / d\Omega$ of $p\bar p \to \omega\eta_c$ and $p\bar p \to \omega J/\psi$. Here, the results are given by taking the range of $E_{cm}=3.8-6.0$ GeV or $4.0-6.0$ GeV with step of 0.2 GeV for the $\eta_c$ or $J/\psi$ production, respectively. The diagrams in the first and third rows are the results without FF while the remaining diagrams are the results with FF, where we take $\Lambda=1.9$ GeV. The diagrams with the same coupling constants $(g_\omega,\kappa_\omega)$ are listed in the same column. The results at the lower and upper limits of $E_{cm}$ range are highlighted with dotted green and short-dashed red lines respectively, while long-dashed blue lines are the result at $E_{cm}=5$ GeV.}
\label{fig:omegaAD}
\end{figure}
\end{center}

In the following, we illustrate the results of $p\bar p \to \omega \Psi$. Just shown in Table. \ref{tab:omega_couplings},
there exist different values of $g_\omega$ and $\kappa_\omega$ for the $pp\omega$ coupling. With $p\bar p \to \omega \eta_c$ and $p\bar p \to \omega J/\psi$ as example, we list the variation of the total cross section of these two processes with $E_{cm}$ when taking three typical combinations of $g_\omega$ and $\kappa_\omega$. Our calculation indicates that the total cross sections of $p\bar p \to \omega \eta_c$ and $p\bar p \to \omega J/\psi$
are dependent on the value of $(g_\omega,\kappa_\omega)$ whether we consider the FF contribution or not (see Fig. \ref{fig:omegaCS_CC} (a)-(d) for more details). The results shown in Fig. \ref{fig:omegaCS_CC} (a)-(d) indeed indicate that the $p\bar p \to \omega \Psi$ processes can be applied to test the values of $(g_\omega,\kappa_\omega)$ listed in Table. \ref{tab:omega_couplings}. It is obvious that the total cross sections of $p\bar p \to \omega \eta_c$ and $p\bar p \to \omega J/\psi$ are suppressed by FF.
Since we cannot constrain the $\Lambda$ value, in Fig. \ref{fig:omegaCS_CC} (e)-(f) we discuss the dependence of the cross section of $p\bar p \to \omega \eta_c$ and $p\bar p \to \omega J/\psi$ on $\Lambda$. For other charmonium productions with $\omega$ meson, the behavior of their cross sections is similar to that of $p\bar p \to \omega \eta_c$ and $p\bar p \to \omega J/\psi$.

In Fig. \ref{fig:omegaCStotal}, we further list the total cross sections for the $p \bar p \to \omega\Psi$
processes taking typical values $\Lambda=1.9$ GeV and $(g_\omega, \kappa_\omega)=(12.2, -0.12)$, where we take the same cutoff $\Lambda$ as that of $p\bar{p}\to \pi^0 J/\psi$. Our calculation shows that the total cross section of
$p\bar{p}\to \omega\eta_c$ is the largest one among all charmonium productions discussed here. The production cross section of $J/\psi$ is roughly 50 times smaller than that of $\eta_c$, while the production cross section of $\chi_{c0}$ is 5 times larger than that of $J/\psi$. We also notice that the total cross sections of $p\bar{p}\to \omega\Psi$ become stable after reaching up to their maximums with increasing $E_{cm}$, which is different from the situation of $p\bar{p}\to \pi^0\Psi$ discussed above.
Just because of the considerable cross sections of $\eta_c$ and $\chi_{c0}$ productions, $p\bar p \rightarrow \omega\eta_c$ and $p\bar p \rightarrow \omega\chi_{c0}$ can be as the ideal channels to study the charmonium production at $\overline{\mbox{P}}$ANDA.

Additionally, with $p\bar p \to \omega\eta_c$ and $p\bar p \to \omega J/\psi$ as example, we also give their center-of-mass frame angular distributions, which are shown in Fig. \ref{fig:omegaAD}. There exists obvious difference among the results calculated with and without FF. We can do a cross check for our calculation, where we take $\Lambda\to \infty$, which denotes that the $pp\omega$ and $pp\Psi$ interactions are treated as the point-like vertices. We find that the result with FF is gradually consistent with that without FF when $\Lambda$ tends to infinity. We also find the node in the differential cross section with FF when taking $\theta=90^\circ$ in the center-of-mass frame, which is similar to $p\bar p\to \pi^0\Psi$. The investigation of $p\bar p \to \omega\Psi$ further shows that the FF contribution cannot be ignored
in the calculation.

\subsection{Total cross sections for other processes }

We can easily extend the formulae of $p\bar{p}\to \pi^0\Psi$ and $p\bar{p}\to \omega\Psi$ to study
$p\bar{p}\to \eta\Psi$ and $p\bar{p}\to \rho\Psi$ respectively, where we only need to replace the corresponding coupling constants and masses. The coupling constant of the $pp\eta$ interaction is taken as $g_{p p\eta}=11.5$ \cite{Downum:2006re}. In Ref. \cite{Machleidt:1987hj}, the $p p\rho$ coupling constants $(g_\rho, \kappa_\rho)=(3.249, 6.1)$ obtained by the Bonn full model. In addition, the masses of $\eta$ and $\rho$ mesons are $M_\eta=547.9$ MeV and $M_\rho=775.5$ GeV \cite{Nakamura:2010zzi}, respectively.

\begin{center}
\begin{figure}[htb]
\scalebox{0.55}{\includegraphics{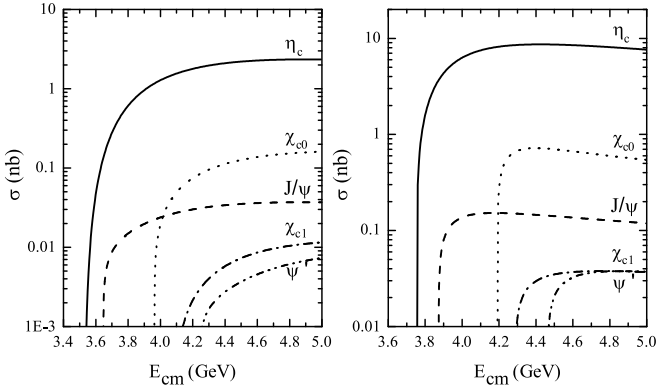}}
\caption{The total cross sections of the processes $p \bar p \to \eta\Psi$ (left-hand) and $p \bar p \to \rho\Psi$ (right-hand) dependent on $E_{cm}$. Here, we take the typical value $\Lambda=1.9$ GeV. }
\label{fig:rhoCStotal}
\end{figure}
\end{center}

In Fig. \ref{fig:rhoCStotal}, the total cross sections of $p\bar{p}\to \eta\Psi$ and $p\bar{p}\to \rho\Psi$ are calculated by including the FF contribution, where we take the typical cutoff $\Lambda=1.9$ GeV. The result also shows that the total cross section of the $\eta_c$ production is the largest one among all charmonium  productions by the $p\bar{p}\to \eta\Psi$ or $p\bar{p}\to \rho\Psi$ processes.

\section{Discussions and conclusion}\label{sec4}

As one of the main physical aims of $\overline{\mbox{P}}$ANDA, studying the charmonium production at the low energy $p\bar{p}$ interaction is an interesting research topic not only for experimentalists but also for theorists.
For investigating the production of charmonium plus a pion by the $p\bar{p}$ annihilation,
the production mechanism shown in Fig. \ref{fig:FeynmanDiagram} was proposed in Ref. \cite{Gaillard:1982zm}, where the $p\bar{p}\to \pi^0 J/\psi$ cross section was calculated. According to this mechanism, the authors in Refs. \cite{Lundborg:2005am,Barnes:2006ck,Barnes:2007ub,Barnes:2010yb} further developed the model of the production of charmonium plus a light meson and applied it to calculate the $p\bar{p}\to \pi^0 \Psi$ processes, where the total cross sections and the unpolarized angular distributions were obtained. Although there are many theoretical groups dedicated to the production of charmonium plus a pion by the $p\bar{p}$ interaction,
further theoretical study of these processes is still valuable, where several further developments of the model were indicated in Ref. \cite{Barnes:2010yb}.

Considering the present research status of the production of charmonium plus a light meson by the low energy $p\bar{p}$ interaction, in this work we revisit this interesting research topic. Different from the former work in Ref. \cite{Barnes:2006ck}, in our calculation we consider the FF contribution to each interaction vertex, which was proposed as one of the possible developments of the model \cite{Barnes:2006ck}. We calculate the total cross sections and the center-of-mass frame angular distributions of the $p\bar{p}\to \pi^0\Psi$ processes. Since the E760 experiment measured the cross section of $p\bar{p}\to \pi^0 J/\psi$ \cite{Armstrong:1992ae}, we compare our result with the E760 data, which indicates that the calculated total cross section of $p\bar{p}\to \pi^0\Psi$ overlaps with the experimental data, where the cutoff $\Lambda$ in FF is taken as 1.9 GeV. The inconsistence between the experimental data and the result without FF in Ref. \cite{Barnes:2006ck} are alleviated by considering the FF contribution, which indicates that the FF involved in each interaction vertex cannot be ignored. Adopting the same $\Lambda$ value, in this work we also give other charmonium production cross sections and the corresponding center-of-mass frame angular distributions of $p\bar{p}\to \pi^0\Psi$. The difference of the results with and without FF further shows that we should consider the FF contribution to our calculation. Thus, these results obtained with FF can be served as further experimental investigation of charmonium production at $\overline{\mbox{P}}$ANDA.

Besides studying the charmonium production plus a pion, in this work we also calculate the $p\bar{p}\to \omega\Psi$ processes. The results of the total cross section and the center-of-mass frame angular distribution of $p\bar{p}\to \omega\Psi$ also indicate the distinct difference of the calculations with and without FF, which is similar to the situation of $p\bar{p}\to \pi^0\Psi$. Different from the $pp\pi$ coupling, the $pp\omega$ interaction is related to two independent coupling constants $g_\omega$ and $\kappa_\omega$, which make $p\bar{p}\to \omega\Psi$ an ideal channel to test different theoretical values of $g_\omega$ and $\kappa_\omega$, where our results also show the total cross sections of $p\bar{p}\to \omega\Psi$ are dependent on the values of $g_\omega$ and $\kappa_\omega$.

In order to reflect the completeness of the study of the charmonium production plus a light meson, we also apply the formulae of $p\bar{p}\to \pi^0 \Psi$ and $p\bar{p}\to \omega \Psi$ to calculate the $p\bar{p}\to \eta \Psi$ and $p\bar{p}\to \rho \Psi$, respectively. The predicted total cross sections of $p\bar{p}\to \eta \Psi$ and $p\bar{p}\to \rho \Psi$ are accessible at the forthcoming $\overline{\mbox{P}}$ANDA.

By the systematic investigation of the charmonium production plus a light meson at the low energy $p\bar{p}$ interaction, we notice a common behavior of the charamonium production, {\it i.e.}, the $\eta_c$ production cross section is the largest one among all discussed processes. This fact shows that the low energy $p\bar{p}$ interaction is an ideal platform to produce $\eta_c$. In addition, we also find that the charmonium production cross sections satisfy the relation $\sigma_{\eta_c}>\sigma_{\chi_{c0}}>\sigma_{J/\psi}>\sigma_{\chi_{c1}}>\sigma_{\psi^\prime}$, which does not depend on the associated light mesons.

\begin{table}[htb]
\begin{center}
\caption{The comparison of the total cross sections of $p\bar p \to m J/\psi (\psi^\prime)$ obtained in this work (the second column) and given in Ref. \cite{Lundborg:2005am} (the third column) and the corresponding $E_{cm}$ value. Here, our results are the typical values of the cross section with FF when taking $\Lambda=1.9$ GeV.}
\label{tab:CStotal}
\begin{tabular}{ccccc}
\toprule[1pt]
~~~~~Reaction~~~~~ & ~~$\sigma_{FF}$ (pb)~~ & ~~$\sigma_{CA}^{max}$ (pb) \cite{Lundborg:2005am}~~  & $E_{cm}$ (GeV) \\
\hline
$p\bar p \to \pi^0 J/\psi$ & 116 & $420\pm40$ & 4.28 \\

$p\bar p \to \eta J/\psi$ & 36 & $1520\pm140$ & 4.57 \\

$p\bar p \to \omega J/\psi$ & 156 & $1900\pm400$ & 4.80 \\

$p\bar p \to \rho J/\psi$ & 127 & $<450$ & 4.80 \\

$p\bar p \to \pi^0\psi'$ & 28 & $55\pm8$ & 5.14  \\

$p\bar p \to \eta\psi'$ & 9 & $33\pm8$ & 5.38  \\

$p\bar p \to \omega\psi'$ & 40 & $46\pm22$ & 5.60   \\

$p\bar p \to \rho\psi'$ & 32 & $38\pm17$ & 5.59   \\
\bottomrule[1pt]
\end{tabular}
\end{center}
\end{table}

The authors in Ref. \cite{Lundborg:2005am} calculated the total cross section of the $J/\psi$ and $\psi^\prime$ productions  $p\bar{p}\to m J/\psi(\psi^\prime)$ by relating these processes with $J/\psi(\psi^\prime)\to m p\bar{p}$. Thus, we also make a comparison of the results obtained by us and those listed in Ref. \cite{Lundborg:2005am}. As shown in Table. \ref{tab:CStotal}, our result with FF of $p\bar{p}\to \omega\psi^\prime, \rho\psi^\prime$ are consistent with those given in Ref. \cite{Lundborg:2005am} while the cross sections of
 $p\bar{p}\to \pi^0\psi^\prime,\eta\psi^\prime$ calculated by us are slightly smaller than those in Ref. \cite{Lundborg:2005am}. The cross section of $p\bar{p}\to \rho J/\psi$ shown in this work also falls into the range of the predicted cross section of $p\bar{p}\to \rho J/\psi$, where the upper limit of the cross section of $p\bar{p}\to \rho J/\psi$ was given in Ref. \cite{Lundborg:2005am}. There also exist the differences of the results of
 $p\bar{p}\to \pi^0 J/\psi, \eta J/\psi, \omega J/\psi$ from this work and Ref. \cite{Lundborg:2005am}, {\it i.e.}, total cross sections presented here are about 4 times, 42 times and 12 times smaller than those for $p\bar{p}\to \pi^0 J/\psi, \eta J/\psi, \omega J/\psi$, respectively.

In summary, in this work we systematically study the charmonium production plus a light meson in the $p\bar{p}$
interaction, and predict the corresponding total cross section and the center-of-mass frame unpolarized angular distribution,
which provide valuable information to the experimental investigation of this kind of reaction. As an ideal experiment for studying the charmonium production, the forthcoming $\overline{\mbox{P}}$ANDA experiment can directly verify the prediction given in this work and test the charmonium production mechanism at the low energy $p\bar{p}$ interaction adopted here.

\section*{Acknowledgement}

This project is supported by the National Natural Science Foundation of
China under Grant Nos. 11175073, 11035006, 10925526,
the Ministry of Education of China (FANEDD under Grant No. 200924,
DPFIHE under Grant No. 20090211120029, NCET, the Fundamental
Research Funds for the Central Universities and the Fok Ying Tung Education Foundation (No. 131006).

\vfill
\newpage

\appendix

\section{The differential cross section of $p\bar{p}\to \omega\Psi$}\label{Appendix_A}

By Eq. (\ref{eq:dt}), we can easily deduce the expressions of the unpolarized differential cross sections, {\it i.e.},
\begin{widetext}
\begin{eqnarray}
    \left\langle \frac{d\sigma}{dt} \right\rangle_{p\bar p \to \omega\eta_c}
    &=& \pi \frac{\alpha_{\omega}\alpha_{\eta_c}}{s(s-4 m_p^2)x^2 y^2}\biggl( \Bigl(-R^2(r^2+2)f^2+(f^2+2R^4+2r^2 R^2+2R^2 f)x y \Bigr) + \kappa \Bigl(3r^2 R^2 f^2- (2f^2+4r^2 R^2)x y\Bigr)
    \nonumber \\
    && - \frac{1}{8} \kappa^2 \Bigl((r^2+8)r^2 R^2 f^2 - (r^2 f^2 + 4f^2- 4r^2 R^4 -4r^2 R^2 f + 16r^2 R^2)x y - 4(R^2+f)x^2 y^2 \Bigr)\biggr),
    \label{eq:etac_diff}
\\
   \left\langle \frac{d\sigma}{dt} \right\rangle_{p\bar p \to \omega\chi_{c0}}
    &=& \pi \frac{\alpha_{\omega}\alpha_{\chi_{c0}}}{s(s-4 m_p^2)x^2 y^2}\biggl(\Bigl(-(r^2+2)(R^2-4)f^2 + \big(2(r^2+2)(r^2-4) + 8r^2 + 2R^4 - 12R^2 + 16 +2(R^2-4)f + f^2 \big)x y\Bigr)
    \nonumber \\
    && + 3\kappa r^2 f \big[(R^2-4)f+2x y \big]- \frac{1}{8} \kappa^2 \big[r^2(r^2+8)(R^2-4)f^2 - \big(4r^4R^2+4r^2R^4-16r^2R^2 +4r^2(R^2-8)f
    \nonumber \\
    && +(r^2-4)f^2 \big)x y - 4(r^2+R^2+f)x^2y^2\big]\biggl),
    \label{eq:chic0_diff}
\\
    \left\langle \frac{d\sigma}{dt} \right\rangle_{p\bar p \to \omega(J/\psi^{(\prime)})}
    &=& \pi \frac{\alpha_{\omega}\alpha_{J/\psi^{(\prime)}}}{s(s-4 m_p^2)x^2 y^2}\Big\{2 \big[-(r^2+2)(R^2+2)f^2 + \big(2(r^2+2)(R^2+2) + 2(r^2+R^2+2)f + 2(r^2+R^2)^2 - 8 + f^2 \big)x y
    \nonumber \\
    && - 2x^2y^2\big] + 2\kappa \big[3r^2(R^2+2)f^2 - (6r^4+14r^2R^2+6r^2f+f^2)x y \big] + \frac{1}{4} \kappa^2 \big[-r^2(r^2+8)(R^2+2)f^2 + (2r^6+6r^4R^2
    \nonumber \\
    && +16r^4+2r^4f +32r^2R^2+16r^2f+r^2f^2)x y + 4(R^2+f)x^2y^2 \big]\Big\},
    \label{eq:psi_diff}
\end{eqnarray}
\begin{eqnarray}
    \left\langle \frac{d\sigma}{dt} \right\rangle_{p\bar p \to \omega\chi_{c1}}
    &=& \pi \frac{\alpha_{\omega}\alpha_{\chi_{c1}}}{s(s-4 m_p^2)x^2 y^2}\Big\{\frac{2}{R^2} \big[-R^2(r^2+2)(R^2-4)f^2 + \big(2R^6+6r^2R^4-8R^4 -16r^2R^2+2r^4R^2+ 2R^2(r^2+R^2-4)f
    \nonumber \\
    && +(R^2+2)f^2 \big)x y - 2R^2x^2y^2 \big] - \frac{2\kappa}{R^2} \big[-3r^2R^2(R^2-4)f^2 + \big(2r^4R^2+6r^2R^4-24r^2R^2 -2r^2R^2f+(R^2+2)f^2 \big)x y
    \nonumber \\
    && + 2(r^2-3R^2)x^2y^2 \big] - \frac{\kappa^2}{4R^2} \big[r^2R^2(r^2+8)(R^2-4)f^2 + \big(2r^6R^2+2r^4R^4-16r^4R^2-24r^2R^4+96r^2R^2+2r^2R^2(r^2
    \nonumber \\
    && +4)f+(2r^2-r^2R^2-8)f^2 \big)x y -2 \big(r^4+r^2R^2+4r^2+2R^4 -12R^2+(r^2+2R^2-4)f \big)x^2y^2 - 4x^3y^3\big]\Big\},
    \label{eq:chic1_diff}
\end{eqnarray}
\end{widetext}
where $\alpha_\omega \equiv g_{p\bar p \omega}^2/4\pi$ , $\alpha_\Psi \equiv g_{p\bar p \Psi}^2/4\pi$ , $r \equiv m_\omega/m_p$ and $R \equiv m_\Psi/m_p$. As the dimensionless variables, $x$ and $y$ are defined as
$x \equiv t/m_p^2-1$ and $y \equiv u/m_p^2-1$, respectively.  $f$ denotes a dimensionless energy variable with definition $f=(s-m_\omega^2-m_\Psi^2)/m_p^2=-(x+y)$.

\section{The total cross section of $p\bar{p}\to \omega\Psi$}\label{Appendix_B}

The detailed formulae of the total cross section of $p\bar{p}\to \omega\Psi$ are
\begin{widetext}
\begin{eqnarray}
    \sigma_{p\bar p \to \omega\eta_c}
    &=& \pi \frac{\alpha_{\omega}\alpha_{\eta_c}m_p^2}{s(s-4 m_p^2)}\Big\{ \big[-R^2(r^2+2)f^2{\cal I}_{2}+(f^2+2R^4+2r^2 R^2+2R^2 f){\cal I}_{1}\big]+ \kappa \big[3r^2 R^2 f^2{\cal I}_{2} - (2f^2+4r^2 R^2){\cal I}_{1}\big]
    \nonumber \\
    && - \frac{1}{8} \kappa^2 \big[(r^2+8)r^2 R^2 f^2 {\cal I}_{2}- (r^2 f^2 + 4f^2 - 4r^2 R^4 -4r^2 R^2 f+ 16r^2 R^2){\cal I}_{1} -4(R^2+f){\cal I}_{0} \big]\Big\},
    \label{eq:etac_total}
\\
   \sigma_{p\bar p \to \omega\chi_{c0}}
    &=& \pi \frac{\alpha_{\omega}\alpha_{\chi_{c0}}m_p^2}{s(s-4 m_p^2)}\Big\{\big[-(r^2+2)(R^2-4)f^2 {\cal I}_{2}+ \big(2(r^2+2)(r^2-4)+ 8r^2+ 2R^4 - 12R^2 + 16 +2(R^2-4)f
    \nonumber \\
    && + f^2 \big){\cal I}_{1} \big]+ 3\kappa r^2 f \big[(R^2-4)f {\cal I}_{2}+ 2{\cal I}_{1} \big] - \frac{1}{8} \kappa^2 \big[r^2(r^2+8)(R^2 -4)f^2{\cal I}_{2} - \big(4R^4R^2+4r^2R^4-16r^2R^2
    \nonumber \\
    && +4r^2(R^2-8)f+(r^2-4)f^2 \big){\cal I}_{1} - 4(r^2+R^2+f){\cal I}_{0}\big]\Big\},
    \label{eq:chic0_total}
\\
    \sigma_{p\bar p \to \omega(J/\psi^{(\prime)})}
    &=& \pi \frac{\alpha_{\omega}\alpha_{(J/\psi^{(\prime)})}m_p^2}{s(s-4 m_p^2)} \Big\{2 \big[-(r^2+2)(R^2+2)f^2 {\cal I}_{2}+ \big(2(r^2+2)(R^2+2) + 2(r^2 +R^2+2)f + 2(r^2+R^2)^2 - 8
    \nonumber \\
    && + f^2 \big){\cal I}_{1}- 2{\cal I}_{0} \big] + 2\kappa \big[3r^2(R^2 +2)f^2 {\cal I}_{2}- (6r^4+14r^2R^2+6r^2f+f^2){\cal I}_{1} \big] + \frac{1}{4} \kappa^2 \big[-r^2(r^2 +8)(R^2+2)f^2 {\cal I}_{2}
    \nonumber \\
    && + (2r^6+6r^4R^2+16r^4+2r^4f +32r^2R^2+16r^2f+r^2f^2){\cal I}_{1} + 4(R^2+f){\cal I}_{0} \big] \Big\},
    \label{eq:psi_total}
\\
    \sigma_{p\bar p \to \omega\chi_{c1}}
    &=& \pi \frac{\alpha_{\omega}\alpha_{\chi_{c1}}m_p^2}{s(s-4 m_p^2)} \Big\{\frac{2}{R^2} \big[-R^2(r^2+2)(R^2-4)f^2 {\cal I}_{2}+ \big(2R^6+6r^2R^4-8R^4 -16r^2R^2+2r^4R^2+ 2R^2(r^2+R^2-4)f
    \nonumber \\
    && +(R^2+2)f^2 \big){\cal I}_{1} - 2R^2{\cal I}_{0} \big] - \frac{2\kappa}{R^2} \big[-3r^2R^2(R^2-4)f^2 {\cal I}_{2}+ \big(2r^4R^2+6r^2R^4 -24r^2R^2-2r^2R^2f+(R^2+2)f^2 \big){\cal I}_{1}
    \nonumber \\
    && + 2(r^2-3R^2){\cal I}_{0}] - \frac{\kappa^2}{4R^2}
    [r^2R^2(r^2+8)(R^2-4)f^2 {\cal I}_{2} + \big(2r^6R^2+2r^4R^4 -16r^4R^2-24r^2R^4+96r^2R^2+2r^2R^2(r^2
    \nonumber \\
    && +4)f+(2r^2 -r^2R^2-8)f^2 \big){\cal I}_{1} -2 \big(r^4+r^2R^2+4r^2+2R^4 -12R^2+(r^2+2R^2-4)f \big){\cal I}_{0} - 4{\cal I}_{-1} \big] \Big\},
    \label{eq:chic1_total}
\end{eqnarray}
\end{widetext}
where $I_m$ ($m=-1,0,1,2$) is defined as ${\cal I}_{m} = \int_{x_0}^{x_1} dx\,{(x y)^{-m}}=\mathcal{I}_m(x_1)-\mathcal{I}_m(x_0)$ with $y = -x - f$,
\begin{eqnarray*}
  x_0 = \frac{ m_{\omega}^2 - 2E_p E_\omega - 2 p_p p_\omega }{m_p^2},\quad
  x_1 = \frac{ m_{\omega}^2 - 2E_p E_\omega + 2 p_p p_\omega }{m_p^2},
\end{eqnarray*}
and
\begin{eqnarray*}
&{\cal I}_{-1}(x) =& -\left(\frac{1}{3}x^3+\frac{f}{2}x^2\right),
\\
&{\cal I}_{0}(x) =& x,\quad {\cal I}_{1}(x) = \frac{1}{f} \ln\left( \frac{x +f}{x} \right),
\\
&{\cal I}_{2}(x) =& \frac{2}{f^3} \ln\left( \frac{x +f}{x} \right) - \frac{1}{f^2} \left( \frac{1}{x +f} + \frac{1}{x} \right).
\end{eqnarray*}
When deducing these expressions of $p\bar{p}\to \omega\Psi$,  the momenta and energies of proton and $\omega$ meson are related to the the Mandelstam variables by
\begin{eqnarray*}
&p_p =& \frac{1}{2}\; \Big(s - 4m_p^2\Big)^{1/2},
\quad
E_p = \frac{s^{1/2}}{2},
\\
&p_\omega =& \frac{1}{2}\sqrt{\frac{(m_\Psi^2-m_{\omega}^2)^2 - 2(m_{\omega}^2+m_\Psi^2)s + s^2}{s}},
\\
&E_k =& \frac{1}{2s^{1/2}}\; \Big(s-m_\Psi^2+m_{\omega}^2\bigg).
\end{eqnarray*}



\begin{thebibliography}{99}

\bibitem{Aubert:1974js}
  J.~J.~Aubert {\it et al.}  [E598 Collaboration],
  Phys.\ Rev.\ Lett.\  {\bf 33}, 1404 (1974).

\bibitem{Augustin:1974xw}
  J.~E.~Augustin {\it et al.}  [SLAC-SP-017 Collaboration],
  Phys.\ Rev.\ Lett.\  {\bf 33}, 1406 (1974).

\bibitem{Nakamura:2010zzi}
  K.~Nakamura {\it et al.}  [Particle Data Group Collaboration],
  J.\ Phys.\ G G {\bf 37}, 075021 (2010).

\bibitem{Lutz:2009ff}
  W.~Erni {\it et al.}  [PANDA Collaboration],
  arXiv:0903.3905 [hep-ex].

\bibitem{Gaillard:1982zm}
  M.~K.~Gaillard, L.~Maiani, R.~Petronzio,
  Phys.\ Lett.\  {\bf B110}, 489 (1982).

\bibitem{Lundborg:2005am}
  A.~Lundborg, T.~Barnes, and U.~Wiedner,
  Phys.\ Rev.\ D {\bf 73}, 096003 (2006).


\bibitem{Armstrong:1992ae}
  T.~A.~Armstrong, D.~Bettoni, V.~Bharadwaj, C.~Biino, G.~Borreani, D.~R.~Broemmelsiek, A.~Buzzo, and R.~Calabrese {\it et al.},
  Phys.\ Rev.\ Lett.\  {\bf 69}, 2337 (1992).

\bibitem{Barnes:2006ck}
  T.~Barnes and X.~Li,
  Phys.\ Rev.\ D {\bf 75}, 054018 (2007).

\bibitem{Barnes:2007ub}
  T.~Barnes, X.~Li, and W.~Roberts,
  Phys.\ Rev.\ D {\bf 77}, 056001 (2008)  [arXiv:0709.4491 [hep-ph]].  

\bibitem{Barnes:2010yb}
  T.~Barnes, X.~Li, and W.~Roberts,
  Phys.\ Rev.\ D {\bf 81}, 034025 (2010).

\bibitem{Downum:2006re}
  C.~Downum, T.~Barnes, J.~R.~Stone, and E.~S.~Swanson,
  Phys.\ Lett.\ B {\bf 638}, 455 (2006).

\bibitem{Cottingham:1973wt}
  W.~N.~Cottingham, M.~Lacombe, B.~Loiseau, J.~M.~Richard, and R.~Vinh Mau,
  Phys.\ Rev.\ D {\bf 8}, 800 (1973).

\bibitem{Lacombe:1980dr}
  M.~Lacombe, B.~Loiseau, J.~M.~Richard, R.~Vinh Mau, J.~Cote, P.~Pires, and R.~de Tourreil,
  Phys.\ Rev.\ C {\bf 21}, 861 (1980).

\bibitem{Nagels:1978sc}
  M.~M.~Nagels, T.~A.~Rijken, and J.~J.~de Swart,
  Phys.\ Rev.\ D {\bf 20}, 1633 (1979).

\bibitem{Machleidt:2000ge}
  R.~Machleidt,
  Phys.\ Rev.\ C {\bf 63}, 024001 (2001).

\bibitem{Sato:1996gk}
  T.~Sato and T.~S.~H.~Lee,
  Phys.\ Rev.\ C {\bf 54}, 2660 (1996).


\bibitem{Mergell:1995bf}
  P.~Mergell, U.~G.~Meissner, and D.~Drechsel,
  Nucl.\ Phys.\ A {\bf 596}, 367 (1996).

\bibitem{Zhu:1999kva}
  S.~L.~Zhu,
  Phys.\ Rev.\ C {\bf 59}, 3455 (1999).

\bibitem{Pire:2005ax}
  B.~Pire and L.~Szymanowski,
  Phys.\ Lett.\ B {\bf 622}, 83 (2005)  [hep-ph/0504255].  


\bibitem{Lansberg:2010mf}
  J.~P.~Lansberg, B.~Pire, and L.~Szymanowski,
  J.\ Phys.\ Conf.\ Ser.\  {\bf 295}, 012090 (2011)  [arXiv:1011.6635 [hep-ph]].  



\bibitem{Lin:1999ve}
  Z.~W.~Lin, C.~M.~Ko, and B.~Zhang,
  Phys.\ Rev.\ C {\bf 61}, 024904 (2000)  [nucl-th/9905003].  





\bibitem{Machleidt:1987hj}
  R.~Machleidt, K.~Holinde, and C.~Elster,
  Phys.\ Rept.\  {\bf 149}, 1 (1987).  



\end{thebibliography}
\end{document}